\documentclass{article}[a4,11pt]
\usepackage{amsmath}
\usepackage{amsfonts}
\usepackage{amssymb}
\usepackage{graphicx}

\begin{document}
	
	\title{Inhomogeneous Evolution of a Dense Ensemble\\
		 of Optically Pumped Excitons\\
		 to a Charge Transfer State}
		\author{Natasha Kirova $^{1,2}$*{$^{,\dagger}$} \\
				$^{1}$ {CNRS,  Laboratoire de physique des Solides},\\
			University Paris-Saclay, 91405 Orsay, France \\
			$^{2}$ Russian Quantum Center, Skolkovo, Moscow 143025, Russia\\
			and \\ 
			Serguei Brazovskii {$^{3\dagger}$}\\
				$^{3}$ {CNRS, Laboratoire de Physique Théorique et Modèles Statistiques},\\
				 University Paris-Saclay, 91405 Orsay, France; brazovsk@gmail.com}
		
	\date{03/05/2025}
	\maketitle
	
	\begin{abstract}
		Phase transformations induced by short optical pulses are mainstream in studies on the dynamics of cooperative electronic states. We present a semi-phenomenological modeling of spacio-temporal effects expected when optical excitons are intricate with the order parameter as in, e.g., organic compounds with neutral-ionic ferroelectric phase transitions.
		A conceptual complication appears here, where both the excitation and the ground state ordering are built from the intermolecular electronic transfer. To describe both thermodynamic and dynamic effects on the same root, we adopt, for
		the phase transition, a view of the Excitonic Insulator---a hypothetical phase of a semiconductor that appears if the exciton energy becomes negative. After the initial pumping pulse, a quasi-condensate of excitons can appear as a macroscopic quantum state that then evolves, while interacting with other degrees of freedom which are prone to an instability.
		The self-trapping of excitons enhances their density, which can locally surpass a critical value to trigger the phase transformation. The system is stratified in domains that evolve through dynamical phase transitions and may persist even after the initiating excitons have recombined.
		
\vspace{12pt}
\end{abstract}

\section{Introduction} \label{Ch.Intro}

There is a persistent interest in phase transformations induced by optical
pulses in cooperative electronic states (see, e.g., a collection of articles
\cite{PIPT book}, proceedings \cite{PIPT-proceedings,IMPACT:2012}, and information from the latest conference \cite{IMPACT:24}). The traditional
technique of pump-induced phase transitions (PIPT) is based on the 
direct excitation between filled and empty electronic bands (the photon absorption
energy $E_{ph}$ is above their separating gap $E_{g}$), creating $e-h$
plasma to affect the correlated electronic states, such as charge density waves
(CDWs), superconductors, or Peierls, Mott, and excitonic insulators. The
shortness of pumps (of the order of femtoseconds) allows one to resolve still disentangled ``fast''
(10--100 fs) electronic and ``slow'' (ps) lattice degrees of freedom. 
 The intensity of the pumps allows for the system to be kicked far away from the equilibrium, which
allows one to follow the dynamical evolution over a wide range of the known phase
diagram; see e.g., \cite{Yusupov:2010}. It becomes possible to also reach
short-leaving ``hidden states'' unavailable in thermodynamic approaches and even
to obtain a truly stable state; see~\cite{Koshihara,Stojchevska:2014}.

The exploitation of generic insulating systems, with their neutral
excitations---the excitons---is still an ill-explored land. The field is
unlimited since many insulators are close to electronic, structural, and/or
magnetic instabilities, and all non-metallic systems possess some excitons
accessible for pumping. The sufficiently dense gas of such excitons may be
ready to form a coherent quantum state close to the Bose--Einstein
condensation (BEC)~\cite{Keldysh:68}. Thus, the PIPT science for insulators
needs to also include studies on the coherent effects in ensembles of excitons.
The BEC of excitons was searched for in bulk semiconductors, particularly in
$Cu_2O$ \cite{CU2O}, and finally observed in bi-layer heterostructures. The BEC of
excitons in the bulk is restricted to a low-temperature $T_{BEC}$ because of
their typically low concentration $n_{ex}$ limited by a "short" $\backsim 10^{-9}$ {s}
life time. This is why the BEC was realized only for much lighter
particles---the polaritons that entangle the exciton and the photon (see
\cite{Berloff:2011} for a review). This limitation was successfully avoided
with a design of artificial bi-layer heterostructures, where electrons
and holes are space-separated; see short reviews in \cite{Shevchenko:2018} or, e.g., \cite{Zhu:95}.

We propose that PIPTs offer another option: under the conditions of femtosecond pumping and
the picoseconds scale of observation, $n_{ex}$ increases to high levels of about
one excitation for a few unit cells, yielding expectations for $T_{BEC}$ up to
$100$~{K}. 
The ``short'' life time becomes more than ``long'' enough.

With good luck, the pumping can go directly to the excitonic level using a
resonant photon absorption $E_{ph}\approx E_{ex}$; then, the excitations appear
initially as delocalized plane waves, since photons create them at zero
momentum $k=0$. In more common cases of a deviation from the resonance
$E_{ph}>E_{ex}$, hot excitons appear, still as delocalized waves but with
a distribution of momenta $k$, so that phonon assistant and/or collision
proOKcesses are required for energy relaxation. Cooling below the bandwidth
of the excitonic spectrum $E_{ex}(k)$ toward its bottom (which is supposed to be at
$k=0$) should lead to the development of a quasi-condensate---the distribution
peak at the lowest energy. The development of the BEC from the initially
incoherent state is an intricate process, having encouraged
many attempts at building a theory at both phenomenological \cite{Berloff:2002}
and microscopic~\cite{Kagan:93,Kagan:97,Gardiner:2002} levels. The resulting picture is not universal as it depends on the interactions of bosons (either pure repulsion for dipole excitons in double-wall structures, or the common for
excitons crossover from attraction at large distances to
repulsion at short ones) and on the inelastic scattering mechanisms required
for cooling.

Almost at the same time that Keldysh suggested the idea of the BEC of
excitons, the concept of an ``excitonic insulator'' (EI) as a special phase of a
semiconductor was introduced~\cite{Kohn,Kohn-2} after a vague proposal in~\cite{Knox:63} and its first theoretical development in~\cite{Cloizeaux}. The
EI became a standard name for a state formed by the condensate of bound
electron--hole pairs, on top of a semiconducting or a semi-metallic state (see
basic reviews~\cite{Halperin,Halperin-2} and more modern ones in
collection \cite{Keldysh75}). The state of the EI appears when the total
energy of the exciton vanishes, $E_{ex}=E_{g}-E_{b}\rightarrow0$. This implies
the possibility of the independent manipulation of the conduction gap $E_{g}$ and
the exciton binding energy $E_{b}$ (e.g., by the composition or the
pressure or the bias voltage). There is some confusion on the terminology since, recently, the name
EI was assigned to states of a kind of CDWs
\cite{EI-exp:TaNiSe,EI-exp:TaNiSe-2,EI-exp:TiSe2,EI-exp:TmSeTe} where the
lattice component is not seen, so the CDW is exclusively  caused by $2P_{F}$
modulations of the electronic density; there can be no excitons with well-screened Coulomb interactions.

In static conditions, the microscopic theory of the thermodynamic EI
state and the theory of the BEC for optically pumped excitons are closely
related, differing mostly in their monitoring parameters: the chemical potential
$\mu_{ex}$ and the density $n_{ex}$ of excitons, respectively. For continuous optical pumping, this duality was strongly emphasized in \cite{Hannewald}.

A dynamical theory of PIPTs is challenging for ab initio microscopic
approaches; see, e.g., \cite{Yonemitsu:2012,Ishihara,Werner}. Nevertheless,
evolution at longer time scales is controlled with collective variables, such as order parameter, the excitons' concentration, and lattice displacements. Then, a
phenomenological approach grants us the possibility to describe oscillatory and
inhomogeneous regimes of dynamical symmetry breaking, which has already been proved to be
in good accordance with experimentation (see, e.g., \cite{Yusupov:2010}).
 Another example was the modeling \cite{Brazovskii:2014-JSNM} of a
 stable hidden state of a polaronic Mott insulator
observed in $1T-TaS_{2}$ \cite{Stojchevska:2014}. The phenomenological approach becomes inevitable
when we consider time dependent spacially inhomogeneous regimes which finally
emerge in modeling.

In this article we study the interplay among the pumping to excitons and the
developing thermodynamic instabilities resulting in an induced phase
transition. We present the phenomenological model which is based on a
generalized concept of the EI formation evolving gradually from the BEC of
excitons. On this bases we perform the modeling of spacio--temporal effects
expected for the optically pumped ensemble of excitons coupled to the order
parameter and lattice displacements. We shall briefly consider the case of
extrinsic excitons which nature is essentially different with respect to the
thermodynamical order parameter; the presence of such  excitons just shifts the
transition temperature and the equilibrium value of the order parameters.
Mainly we shall be concerned with a conceptually most interesting case of intrinsic
excitons which density contributes directly to the thermodynamic order
parameter; this is the case of the charge transfer excitons with respect to
the charge ordering transition. Keeping in mind the already existing
experiments involving neutron-ionic ferroelectric transitions, we shall include also possible
interactions with another degree of freedom---the lattice displacements.

We shall report on two major observations. 1. The interaction of excitons with
the order parameter can lead to their self-trapping thus enhancing their local
density which can locally surpass a critical value to trigger the phase
transformation. Then the system is stratified in domains which evolve through
dynamical phase transitions and may persist even after recombination of pumped
excitons. 2. The excitons (unlike well studied cold Bose atoms) allow for
virtual processes of creation/annihilation of their pairs from/to the vacuum
which lead to a dynamical transition of the phase locking, thus indicating the
demarcation between regimes of BEC and the EI. 

We model some effects
of transformations among the condensed and incoherent reservoirs of excitons.
First, we shall study stochastic equations emulating a random falling of hot
particles to the condensate. Second, we shall model a two-fluid macroscopic
coexistence with a gradual absorption of the initially pumped particles to the
condensate. Preliminary results and discussions can be found in earlier
authors' publications
\cite{SB+NK:2014,SB+NK:2015,SB+NK:2016-prb}.

The article is organized as follows: Section \ref{Ch.IM-EX} is alone devoted
to the case when the excitons are foreign with respect to the order parameter. In Section
\ref{Ch.Intr-ex} we introduce the intricate case of the EI and the BEC. In Section
\ref{Ch.EI-MF} we present multi-field model for the  EI complemented by lattice
displacements. Section \ref{Ch.Incoh} is dedicated to the effects coming from
initially present hot excitons. Section \ref{Ch.Con} is devoted to conclusions.

\section{Extrinsic Excitons and an Arbitrary Order Parameter}
\label{Ch.IM-EX}

The conceptually simple and expectedly most common situation refers to the
case of extrinsic excitons which nature is different with respect to that of
the order parameters $\eta$. Their interaction just affects the transition
temperature and the equilibrium value of $\eta$ while variations of $\eta$
shift the exciton energy.

The single exciton is a quantum state in both the internal and the center of mass coordinates. 
Here and throughout the article, our basic assumption will be that, after a high
initial pumping, the quasi-condensate of optically pumped excitons appears
sufficiently early as the macroscopic quantum state. The condensate
can be described by a macroscopic wave function $\Psi=|\Psi|\exp(i\zeta)$
which evolves interacting with other degrees of freedom prone to an instability.
The attempt to describe a coexistence of the BEC of excitons and their
initially created incoherent liquid will be given below in Section \ref{Ch.Incoh}.

Suggesting for simplicity the smallness of $\eta$ in a proximity to the
(nearly) second order phase transition from the high temperature (T) state
with $\eta=0$, we can write the energy functional density for $\eta$ and
$\Psi$ as

\begin{align}
W(\eta,\Psi)= & W(\eta)+(E_{ex}^{0}+
g_{1}\eta+g_{2}\frac{\eta^{2}}{2})|\Psi|^{2}+\frac{k}{2}|\Psi|^{4}
+\frac{\hbar^{2}}{2M}\left(\nabla\Psi\right)^{2}
\label{W-eta-psi} \\
& W(\eta)  =\frac{a}{2}\eta^{2}+\frac{g_{3}}{3}\eta^{3}+\frac{b}{4}\eta^{4}+\frac{c}{2}\left(\nabla\eta\right)^{2}
\nonumber
\end{align}

{Here the} odd in $\eta$ terms (the ones with coefficients $g_{1,3}$) are
present if the order parameter is not symmetry breaking which is commonly a
case of the I$^{\mathrm{st}}$ order transition close to the II$^{\mathrm{nd}}$
order one (the critical point of monitoring parameters where $a=0$). The odd
$\varpropto g_{3}\eta^{3}$ term with $g_{3}<0$ can give rise to the
second, above the first one at $\eta_{1}=0$, minimum at some $\eta_{2}>0$
bringing to life the metastable state which further on becomes the
preferable one. At presence of excitons with a density $n_{ex}=|\Psi|^{2}$,
their interaction with $\eta$ shifts the critical point from $a_{cr}=0$ to
$a_{cr}^{^{\prime}}=-g_{2}n_{ex}$. Reciprocally, the appearance of $\eta\neq0$
shifts the exciton energy from $E_{ex}^{0}$ to $E_{ex}^{0}+g_{2}\eta^{2}/2$.
That can be verified by observing the increasing value of the
exciton level in the ordered state $\eta\neq0$. This situation is not a dogma
and the opposite sign $g_{2}<0$ may take place. Intuitively, the sign of
$g_{2}$ is expected to be positive if $\eta$ drives the system towards further
dielectrization and negative if the system is driven towards metallization.

The dynamics of this system can be studied by the set of equations
\begin{align}
i\hbar\partial_{t}\Psi+i\hbar\Gamma\Psi &  =\frac{\delta W(\eta,\Psi)}{\delta\Psi^{\ast}}\label{t-psi-eta}\\
\frac{a}{\omega^{2}}\frac{\partial^{2}\eta}{\partial t^{2}}+\gamma
\frac{\partial\eta}{\partial t} &  =-\frac{\delta W(\eta,\Psi)}{\delta\eta}\label{t-eta-psi}\\
\int|\Psi|^{2}dV &  =N=\bar{n}_{ex}V~~
\end{align}
{Here}
 $W$ is given by Equation (\ref{W-eta-psi}), $\omega$ is the bare frequency of
oscillations of a weakly perturbed field $\eta$, $\gamma$ is their attenuation
parameter, $\Gamma$ is the decay rate in the ensemble of excitons, $V$ is the
system volume. Above, we have ignored effects breaking the conservation of the
total number $N$ of excitons which will be important later for the case of
intrinsic excitons, in Section \ref{Ch.Intr-ex}.

The traditional and still most common practice in PIPT, at least with conventional mechanisms of e-h excitations, is that the pumping is applied to a system below (i.e., $a<0$) a second order (i.e., at $g_1=g_3=0$ ) phase transition. 
It destroys temporarily the low T state with  $\eta=\pm\eta_0=\pm\sqrt{(-a/b)}\ne0$ by increasing the efficient $a$ from the initial negative
value $a<0$ to the positive one $a^{\prime}=a+g_{2}n_{ex}>0$, hence expecting $g_{2}>0$. The new equilibrium state is displaced almost instantaneously to $\eta=0$ while the system is still prepared at the state with $\eta=\eta_0$. Such a strongly non-equilibrium situation is resolved in pendulum oscillations spanning the large interval  within $-\eta_0<\eta<\eta_0$. Because of the attenuation $\sim$$\gamma$ the amplitude of oscillations keeps squeezing towards the new equilibrium at $\eta=0$. But well earlier, the recombination of excitations reduces their concentration $n_{ex}$ to a critical value when $a^{\prime}$ changes back the sign and the new pair of minima at $\pm\eta_0^{\prime}\ne 0$ is worked out. Finally the pendulum oscillations are trapped in one of these minima which values gradually return to the bare ones at $n_{ex}=0$. This scenario was recovered in experiments as e.g., in \cite{Yusupov:2010}. 

With the later PIPT tendency of searching for ``hidden states'', the preferable scenario became to reach the
second (meta)stable state at some $\eta_{2}\neq0$. Its existence  is imperative for
non-symmetry-breaking transitions where $g_3\ne 0$. That does not exclude some rare symmetry breaking cases where the negative higher order
therms, like the coefficient $b$, are known to occur, leading also to a weakly Ist order phase transition. 
In these cases, at high T the lowest state 1 with $\eta=0$ is
stable and the higher energy state 2 with $\eta=\eta_{2}$ is metastable; the states are separated by a barrier at some $\eta=\eta_{b}$. 
With decreasing $a$ the binodal is reached when both states 1 and 2 are of the same energy. If the state 1 survives as now the metastable state, than a spinodal point $a=0$, with no more barrier, can be reached in similarity to the above case of the IInd order transition.

Now two directions of PIPT are envisionable with different types of switching: i. from the high T state 1 to the state 2 provided that $g_2<0$ or ii. from the low T state 2 to the state 1 provided that $g_2>0$. The numerical modeling of a kind of such a system 
can be found in \cite{SB+NK:2014}. An intrigue was that in a certain interval around the barrier $\eta_b$, the energy as a function of $\eta$ has a negative curvature bringing to life the space homogeneous regime with rich patterns of arising, colliding, and merging domains and their walls. Beyond the conventional in thermodynamics picture of the separation of coexisting phases, in PIPT dynamics these patterns serve also as self-trapping areas for excitons where their local density is enhanced with respect to the mean one, then the induced transitions take place locally even before the mean value of the critical pumping is reached for the whole volume.

\section{Intrinsic Excitons and the Excitonic Insulator}
\label{Ch.Intr-ex}

\subsection{Generic Excitonic Insulator}

Now we turn to the conceptually more complicated case where the density of
excitons and the order parameter, with a magnitude $\rho$, are of the same origin; they are
additive and statically indistinguishable. This dualism is illustrated by
physics of phase transitions from a (quasi)neutral to ionic states in
donor-acceptor molecular stacks (see \cite{crystals:17} for the latest review
and the analysis of the phase diagram, \cite{Koshihara,Okamoto-rev} for a
review of PIPT experiments, \cite{Okamoto:12} for their later development, and
our relevant discussion in \cite{SB+NK:2016}). In these compounds, the high-T
state already shows a charge alternation $\rho_{1}$ among donor and acceptor
sites which is natural because of their different affinities. But at lower T
the phase transition takes place to the ``ionic'' state characterized by a
higher $\rho_{2}>\rho_{1}$ (in units of the electron charge $e$, typically
$\rho_{2}\approx0.7,\rho_{1}\approx0.3$). Apparently, this transition is not
symmetry breaking, but commonly also another, symmetry breaking, order
parameter enters the game. This is a spontaneous dimerization of bonds $u$
(akin to (spin)Peierls effect) on top of the either built in or spontaneously
augmented dimerization of sites; we shall return to this issue below in
Section~\ref{Ch.EI-MF}.

The charge transfer present in the ground state can be viewed as a condensate
of excitations composed with e-h pairs which otherwise need to be excited
across the excitation gap. The appropriate wave function of the e-h state is
composed with all possible eigenstates including their free conducting states
and bound ones---the Wannier-Mott excitons (called equivalently the ``charge-transfer excitons''). 
But the fortunate fact, that, in
the material of the primary interest, the e-h continuum gap lies three times
higher than the energy $E_{ex}$ of the excitons, allows us to concentrate only
on the later ones. Thus, we arrive at the picture of the Excitonic Insulator
suggested long time ago for more conventional semiconductors. The incoherent
optical pumping firstly gives rise to a gas of uncorrelated excitons 
which density might be additive to the one
frozen in to the ground state. But if the initial excitons have a time to be
cooled down to the BEC state, then both the BE and the EI condensates are
described by the unique wave function $\Psi$. This integrity is demonstrated
by the duality of theoretical descriptions of the BEC of excitons and of the
EI~\cite{Keldysh:73}. The formulations differ only by the monitoring
parameters: the mean density for the BEC versus the chemical potential for the
EI. Microscopically, the interpolation among the BEC and the EI was considered
in \cite{Comte:82,Littlewood:04}.

Since we concentrate upon a situation with the first order transition, then in
equilibrium, without pumping, the energy functional $W(\rho)$ must have two
minima: the stable one at $\rho_{1}$ and the metastable one at $\rho_{2}$
separated by the barrier $W_{b}=W(\rho_{b})$ at $\rho_{b}$. The first order
phase transition happens at the temperature $T_{c}$ when $W(\rho_{1})=W(\rho_{2})$. 
In spite of the essential jump in $\rho$ observed
experimentally, there are clear indications that the first order phase
transition is close to a second order one. This proximity helpfully allows to
use the Landau expansion for the energy functional which (for a homogeneous state) can be written as
\begin{align}
W(\Psi)  &  =W_{\rho}+W_{phase}(\Psi)\label{W-psi}\\
W_{\rho}  &  =E_{ex}^{0}\rho+\frac{a}{2}\rho^{2}+\frac{b}{3}\rho^{3}
~,~\rho=\Psi^{\ast}\Psi+n_{incoh} 
\label{W-rho}
\end{align}
In (\ref{W-rho} the density $n_{incoh}$ of the incoherent cloud of higher energy excitons is
taken into account for further purposes to which we shall return in
Section \ref{Ch.EI-MF}. The term $W_{phase}(\Psi)$ in (\ref{W-psi}) is a special energy contribution depending on the
phase of $\Psi$ which we shall discuss in this section below. 
The energy $W_{\rho}$ depends only on the exciton
density; its derivative gives the thermodynamic definition of the exciton~energy

\begin{equation}
\frac{\partial W_{\rho}}{\partial\rho}=U(\rho)=E_{ex}^{0}+a\rho+b\rho^{2}\label{U}
\end{equation}
The term $W_{phase}(\Psi)$ in (\ref{W-psi}) is a special energy contribution depending on the
phase of $\Psi$ which we shall discuss just below. Notice in (\ref{W-rho}) the
linear dependence of the energy upon the explicit order parameter $\rho$.
Actually the energy starts quadratically in therms of the hidden parameter
$\Psi$. Also, being small, $\rho$ cannot turn negative because $\rho=\Psi^{\ast}\Psi>0$ 
which constraint cannot be seen thermodynamically.

\begin{figure}
\includegraphics[width=7.5cm]{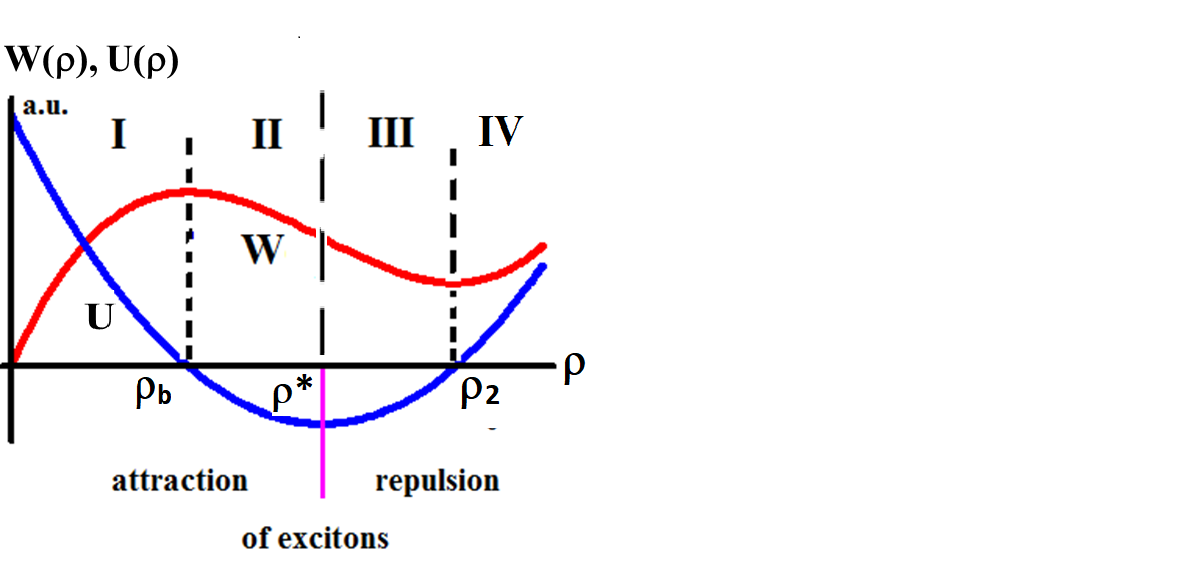}
\caption{ Plots of the energy $W(\rho)$ (red) and its derivative -- the potential $U(\rho)$. 
Four regions of different stability are separated by vertical dashed lines indicating the critical values $\rho_b,\rho^*,\rho_2$.
I and II: decreasing $U(\rho)>0$ -- the attraction of excitons,
II and III: $U(\rho)<0$ -- spontaneous creation of excitons, 
III and IV: increasing $U(\rho)>0$ -- repulsion of excitons.}
\label{fig:1}
\end{figure}

The energy $W(\rho)$ can have two minima: the stable one which we have defined in (\ref{W-rho}) as
$\rho_{1}=0$ and a metastable one at $\rho=\rho_{2}$ separated by the barrier
at $\rho_{b}$. This standard form is significant in terms of the excitons as
showing four different regimes indicated in Figure \ref{fig:1}.
In the region I $W(\rho)$ rows with $\rho$, hence the exciton energy $E_{ex}(\rho)>0$, but it decreases
with increasing $\rho$ which corresponds to the attraction of excitons. In the
regions II,III $E_{ex}\eqslantless0$ indicating the regime of the EI state.
Region II and III are separated by the inflection pnt  $\rho^{\ast}$ of the function
$W(\rho)$ where $E_{ex}<0$ is minimal; at $\rho>\rho^{\ast}$, the E$_{ex}$,
being still negative, starts to grow which indicates on the excitons'
repulsion. Region II and IV are separated by the metastability point where $E_{ex}(\rho_{2})=0$ 
and it keeps growing with increasing of $\rho$ with the feature of the BEC.

Integration of the BEC of excitons and the EI approaches allows for the
combined description for the dynamics of excitons and the evolution of the
phase transition after the pumping. It includes oth thermodynamic and dynamic
effects of the PIPT. Still, adopting the concept of the EI we arrive at the
enigma. The dualism of the excitonic density $|\Psi|^{2}$ and the
thermodynamic order parameter $\rho$ do not seem to be consistent. Within the
thermodynamic picture, the order parameter $\rho$ is a single real
non-conserved field. 
The expected equation (recall the conventional Equation \eqref{t-eta-psi}) allows for unrestricted evolution
towards the energy minimum in $\rho$:
\[
\frac{d^{2}\rho}{dt^{2}}\varpropto-\frac{\delta W}{\delta\rho}.
\]
which was very well verified in PIPT science.

The charge transfer density from excitons $n_{ex}=|\Psi|^{2}$ is also the real
field which seems to be equivalent or additive to $\rho$. Nevertheless, its
evolution is described by the complex wave function $\Psi=|\Psi|\exp(i\zeta)$ of
the BEC. The corresponding evolution of the excitonic condensate contains the
hidden degree of freedom, the phase $\zeta$. The evolution of $\zeta$ is
measurable through the instantaneous energy of the exciton (defined
dynamically as $E_{ex}=-\hbar\partial_{t}\zeta$) which in the homogeneous
regime (or in average over the space) coincides with the thermodynamic
definition $E_{ex}\Rightarrow U={dW}/{d\rho}$.

The excitons are the interacting bosons which at sufficiently high
concentration and low temperature can be described collectively by an adapted
Gross--Pitaevskii theory. This is a generalization of the nonlinear
Schr\"{o}dinger equation (NLSE) which conserves the total number of particles,
hence no evolution of $\rho$ is allowed.

\begin{equation}
i\hbar\frac{\partial\Psi}{\partial t}=
-\frac{\hbar}{2M}\frac{\partial^{2}\Psi}{\partial x^{2}}+U(\rho)\Psi
\ ,~\int\Psi\Psi^{\ast}dx=cnst \label{shr2}
\end{equation}

\begin{figure}
\includegraphics[height=5cm,width=7.5cm]{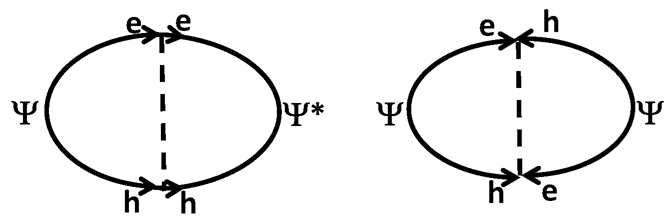}
\caption{{Contributions of normal (at the left diagram) and anomalous (at the right one) Coulomb interactions to the energy of the EI. The dashed line shows the e-h interaction. The semi-circular lines with arrows correspond to anomalous Green functions $\propto \Psi,\Psi^*$ which mix electronic (e) and hole (h) creation/annihilation operators.} 
}
\label{fig:2}
\end{figure}

To make the evolution of the fields $\Psi$ and $\rho$ compatible, we need to
take into account the processes which do not conserve the number of particles,
i.e., given by phase-dependent interactions. They appear from anomalous matrix
element of Coulomb interactions transferring two electrons across the gap,
from filled to empty band \cite{Keldysh:73} which have been studied intensively
for bi-layer systems (\cite{lozovik-jetp} with a recent return of interests in
\cite{Millis:2024}). That means the simultaneous creation or annihilation of
two e-h pairs, i.e., creation/destruction of two excitons from/to the vacuum.
The contributions of responsible interband Coulomb interaction, together with the principle diagonal one, are shown in Figure \ref{fig:2}.
Being virtual usually (the energy is not conserved), these transition
amplitudes acquire nonzero averages for macroscopic concentrations in the BEC
or EI states.

Finally, the EI free energy acquires an additional contribution which fixes the phase, hence violates the particles number conservation.
\begin{equation}
\Delta W_{phase}(\Psi)=\frac{1}{2}(A^{\ast}\Psi^{2}+A\Psi^{\ast2})
=|A|\rho\cos[2(\zeta-\alpha)] 
\label{W-phase}
\end{equation}
({From}
 here on, we take the material constant $\alpha=0$ with $A>0$ which can
always be done by choosing the origin of the variable phase $\zeta$.)

By the definition of the EI, the phase fixing terms are small, $A\ll1$, hence
the total complex order parameter $\Psi$ still needs to be exploited. The
relative value of anomalous $A$ terms, with respect to the main Coulomb energy
$E_{b}$, is of the order of $A/E_{b}\propto(a/R)^{d}$ (where $a$ is the lattice
spacing, $R\gg a$ is the exciton radius, and $d$ is the space dimension) which
is small by definition for Wannier--Mott excitons. For donor-acceptor chains,
the bare value of $A$ may not be very small as $R$ is only of a few $a$ and
$d=1$. But the fact that the system is nearly one--dimensional results in
strong phase fluctuations which reduce functions periodic in $\zeta$. Then the
effective value 
$A\rightarrow A\langle exp(2i\zeta)\rangle\approx A\exp(-2\langle\zeta^{2}\rangle)$ 
can be small and even be renormalized to zero (in the context of excitons in two-dimensional structures, effects of the spacial phase incoherence where discussed in \cite{Shevchenko:2018}).

At a low but still macroscopic density of residual excitons, $\rho\rightarrow0$, $U\rightarrow E_{ex}^{0}$, the wave function oscillates now with two mirror frequencies: $\pm\omega_{A}=\pm E_{A}/\hbar$
where the energy $E_{A}<E_{ex}^{0}$ is shifted down with respect to the bare
$E_{ex}^{0}$. The eigenfunction of the single exciton appears as the interference:
\begin{align}
\Psi &  \propto\sqrt{1-|A|/E_{ex}^{0}}\cos(\omega_{A}t)+
i\sqrt{1+|A|/E_{ex}^{0}}\sin(\omega_{A}t),
\label{psi(tt)}\\
|\Psi|^{2}  &  \propto1-|A|/E_{ex}^{0}\cos(2\omega_{A}t)
\nonumber
\end{align}
{This quantum}
 interference of positive and negative energies leading to
persistent coherent oscillations of the number of excitons was recently
interpreted as a dynamical Josephson effect predicted for excitonic
condensates in double-well heterostructures or natural bi-layer systems
\cite{Millis:2024,Lozovik-pisma}.

\subsection{Evolution Equations for the Generic Excitonic Insulator}
\label{Ch.Ev-Eq}

We shall adopt the Gross--Pitaevskii type equation which is applicable when,
for all relevant states of bosons, their occupation numbers are much bigger
than unity. For excitons on a $d$--dimensional lattice, this condition reads that
their mean density per lattice site is $n_{ex}\gg(T/D)^{d}$. With the
exciton bandwidth $D\sim$$10^{2}\ K$, this inequality can be satisfied
for a typical experimental temperatures $T\sim10^{1}$. Even if the initial
value of $\rho$ is not sufficiently high, the kinetics of cooling feeds the
low energy states, then the Gross--Pitaevskii theory becomes applicable sooner
or later. This is an advantage of the fast PIPT technique, where the time of
observations is shorter than the recombination life time of excitons.

The basic equation can be chosen as
\begin{equation}
i\hbar\partial_{t}\Psi=
-\frac{\hbar^{2}}{2M}\partial_{x}^{2}\Psi+U(|\Psi|^{2})\Psi-A/2\Psi^{\ast}-i\hbar\Gamma\Psi 
\label{psi}
\end{equation}
{Here two}
 terms appear breaking the conservation of total number of excitons:
the nondissipative term $\varpropto A$ and the common dissipative one
$\sim$$\Gamma$. The relaxation rate $\Gamma$ has a sophisticated behavior:
depending on $\rho$, $\Gamma$ evolves from a constant $\Gamma=1/\tau_{rec}$ at
vanishing $\rho$ (the single-particle recombination of rare excitons) to
$\Gamma\propto\rho$ at moderate $\rho$ (Bose--Einstein statistics and also
Auger processes, well proved in semiconductors). Nevertheless, around the
second minimum of $W$ at $(\rho\approx\rho_{2})$ of the high density
equilibrium phase, we expect that $\Gamma(\rho)\rightarrow0$ since there is no
channel for decay. (We neglect the evaporation over the barrier towards small
$\rho=0$.) We can interpolate among these regimes writing $\Gamma=G(\rho)\rho
U(\rho)/\hbar$, where $G(\rho)$ is some structureless dimensionless function
of $\rho$ (which we shall choose as a constant for the numerical modeling). In
the region II, III of Figure \ref{fig:1} $\Gamma<0$ corresponding to the creation of
excitons---the amplification of the number of excitons instead of their decay.
Another way to reason is to notice that the energy relaxation terminates when
$\partial_{t}\zeta=0$ which leads to another form of $\Gamma$, equivalent in a
space homogeneous regime:

\begin{equation}
\Gamma\Rightarrow-\frac{G}{2i}(\Psi^{\ast}\partial_{t}\Psi-\Psi\partial_{t}\Psi^{\ast})
=-G\rho\partial_{t}\zeta=\frac{G(\rho)\rho}{\hbar}E_{ex}(t,x),\ G\approx const. 
\label{Gamma}
\end{equation}

\subsection{Space Homogeneous Regime or Zero Dimension}

Enforcing the space homogeneous regime $\nabla\Psi\equiv0$ or for a a zero-dimensional system, 
the equations for $\rho,\zeta$ acquire a simple form
\begin{align}
\hbar\partial_{t}\zeta &  =-U(\rho)+|A|\cos(2\zeta)\label{dphase-dt}\\
\partial_{t}\rho &  =
-\Gamma\rho+\left\vert A\right\vert \rho\sin(2\zeta)=
G\rho^{2}\partial_{t}\zeta+\left\vert A\right\vert \rho\sin(2\zeta)
\nonumber \\
&  =-(G/\hbar)\rho^{2}(U(\rho)-|A|\cos(2\zeta))+\left\vert A\right\vert\rho\sin(2\zeta).
\label{drho-dt}
\end{align}

The static solution $\partial_{t}\Psi=0$ becomes possible taking into account the phase locking effect when
\[
\sin(2\zeta)=0~,-U(\rho_{st})+|A|\cos(2\zeta)=0
\]
i.e.,
\[
\zeta_{st}=\pi n/2\ ,\ U(\rho_{st})=|A|(-1)^{n}
\]
{The presence}
 of the $A$ term results in shifting the static value $\rho_{st}$
with respect to the thermodynamically equilibrium value $\rho_{eq}$.

The dynamics is illustrated by the numerical solution of the above equations.
For the modeling we specify the ground state energy as
\[
W(\rho)=\rho((\rho-1)^{2}+0.05)/1.05\ ,\ U(\rho)=dW/d\rho
\]
where $W(\rho)$ is normalized to have the bare exciton energy $E_{ex}^{0}=U(0)=1$. With these parameters, we are below the thermodynamic phase transition to the EI state but it can exist as a metastable state (the minimum of $W(\rho)$ at $\rho_{2}\approx1$).

\begin{figure}
	\includegraphics[width=0.45\linewidth]{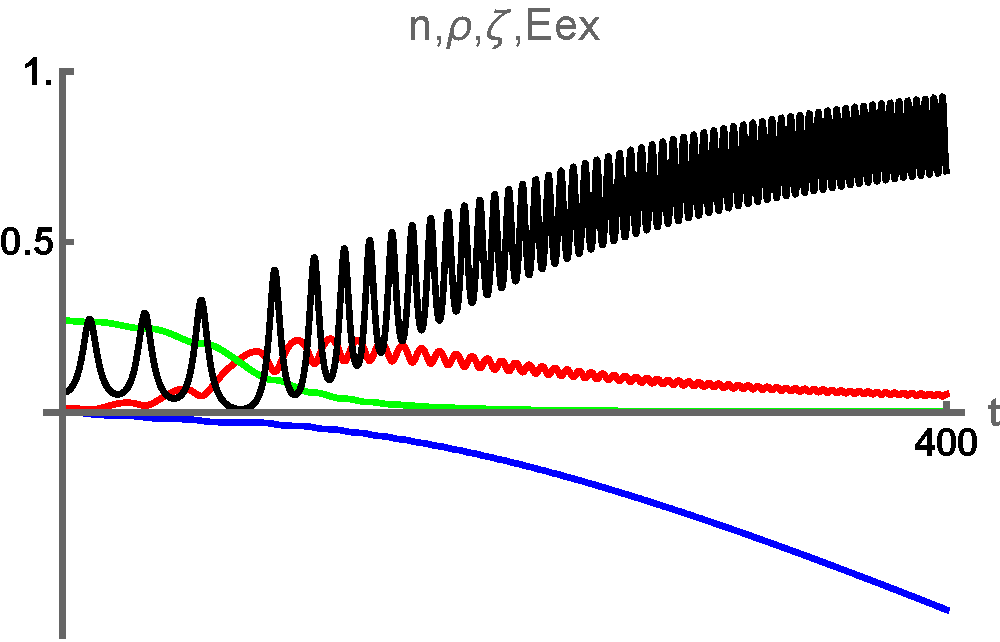}
	\qquad
	\includegraphics[width=0.45\linewidth]{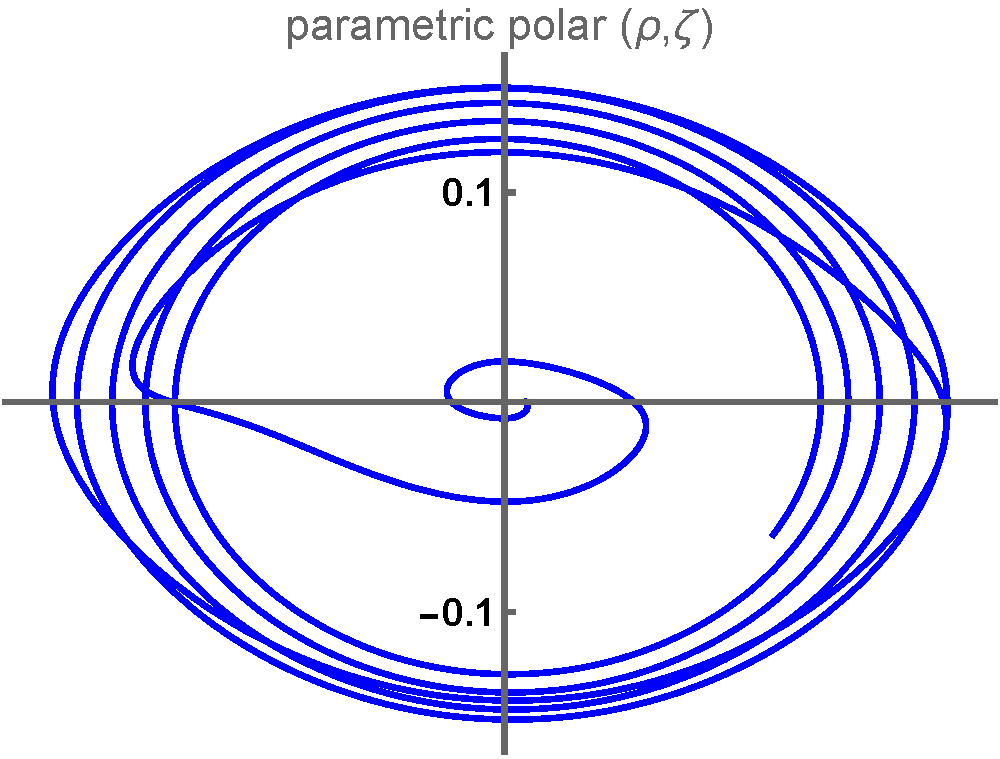}
	\newline
	\includegraphics[width=0.45\linewidth]{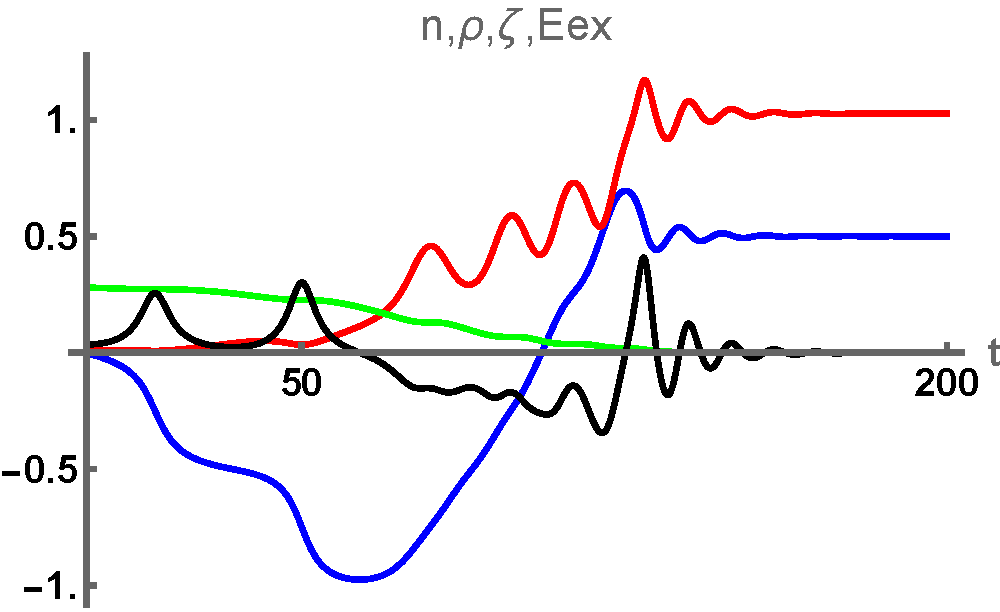}
	\qquad
	\includegraphics[width=0.45\linewidth]{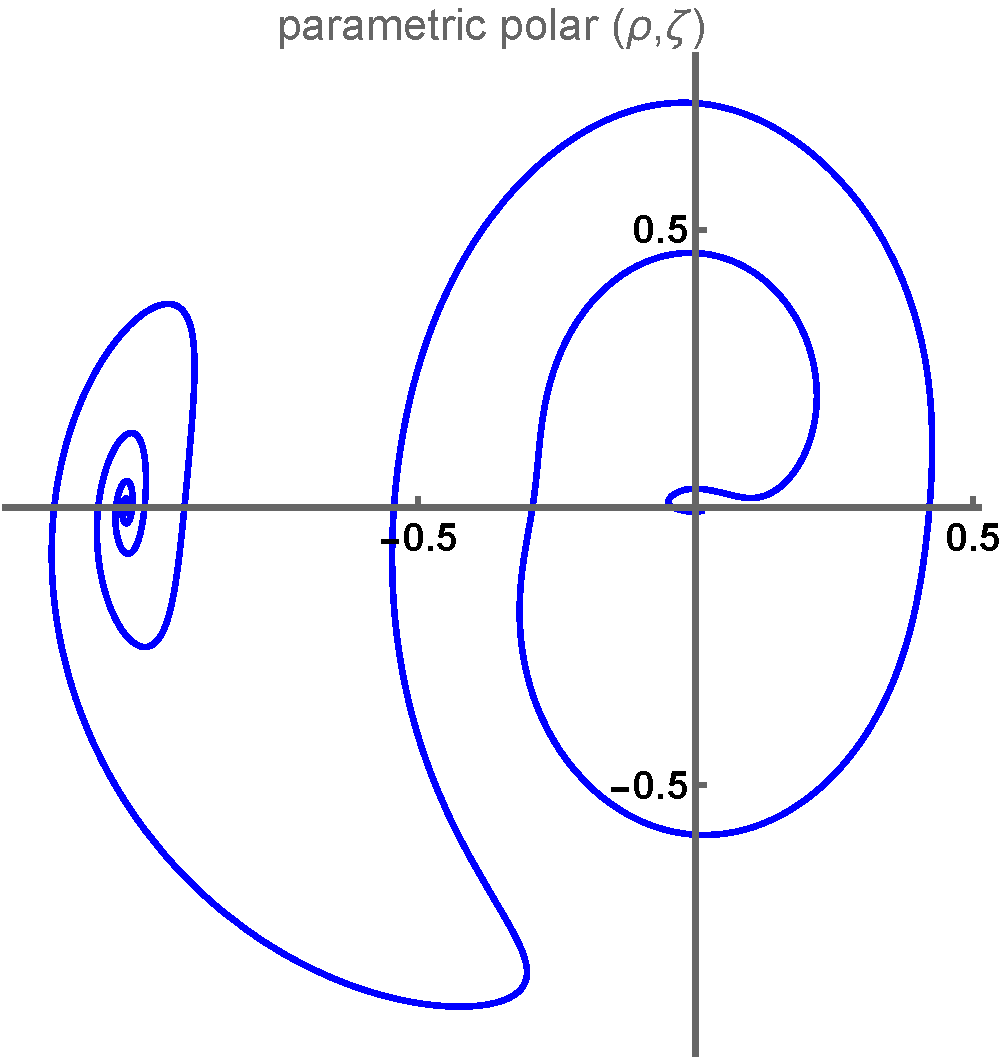}
	\caption { { Upper row: subcritical pumping, lower row: Supercritical pumping. Left panels: plots for the excitons density $\rho(t)$ -- red, the excitons energy $E_{ex}(t)$ -- black, the phase $\zeta(t)/\pi$ -- blue, and the distributed pumping profile $n(t)$ -- green. Right panels: Parametric polar plots $(\rho,\zeta)$.}}
	\label{fig:3}
\end{figure}

Figure \ref{fig:3} shows time dependencies of the amplitude, the phase, the exciton energy, and also the
trajectory for $\rho(\zeta)$. 
The sub-barrier pumping leaves the system in the unlocked regime and it relaxes to the virgin
ground state with $\rho\rightarrow0$ and $E_{ex}(t)\rightarrow E_{ex}^{0}$ with linearly growing $-\zeta$. 
In the super-barrier regime, $\rho$ grows reaching the locking
transition where the exciton energy $E_{ex}(t)\rightarrow E_{loc}$ gradually
freezes at the thermodynamic equilibrium while the phase is locked via the
dynamical transition at an allowed value $\zeta=\pi(n+1/2)$. For both cases the
curves are superimposed by oscillations originated via the $A$ term by
destruction/creation of pairs of excitons. 
 The calculations presented in Figure \ref{fig:3} have been actually performed within a more general scheme taking into account a distributed pumping (the green line in the plots); we shall return to these figures in  Sec.\ref{2fluids}.

\subsection{Inhomogeneous Regimes and Space-Time Patterns}

\begin{figure}
\includegraphics[width=0.3\linewidth]{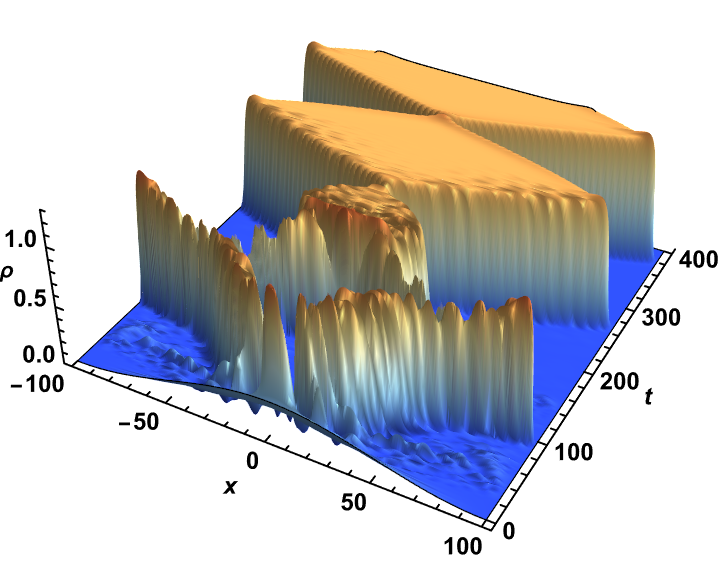}
\quad
\includegraphics[width=0.3\linewidth]{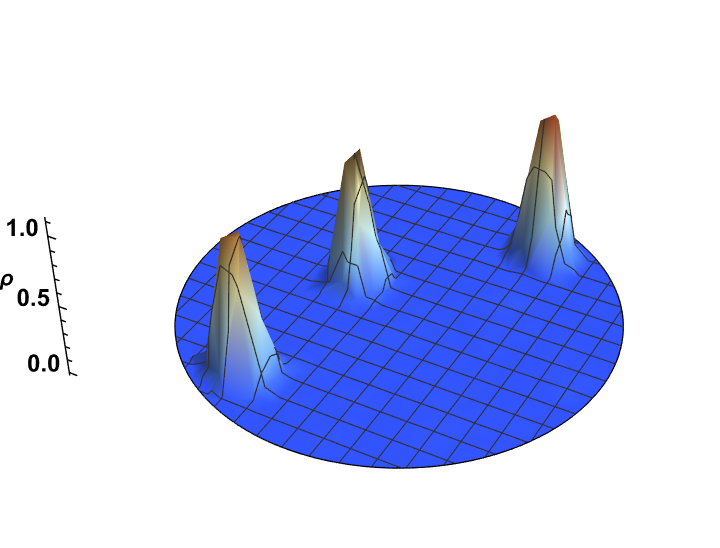}
\quad
\includegraphics[width=0.3\linewidth]{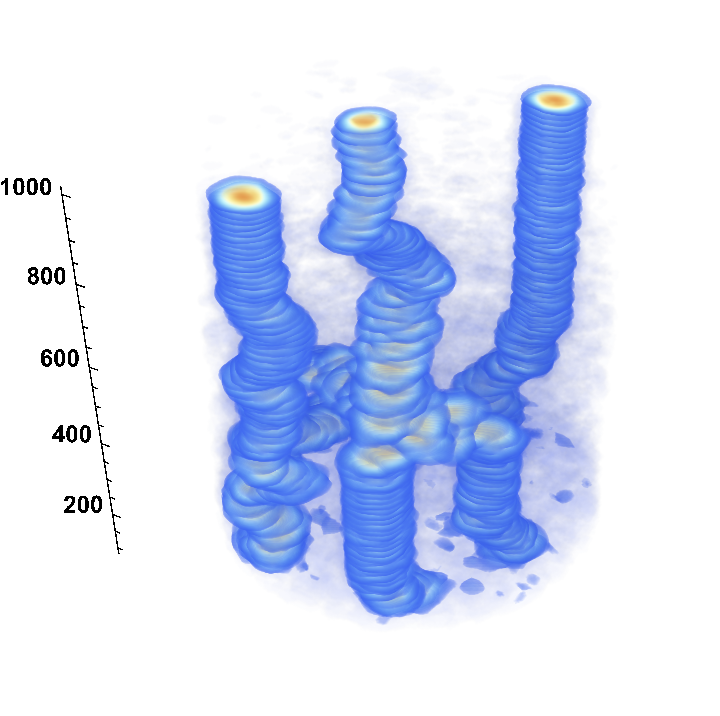}
\caption{
Plots for an inhomogeneous evolution.
	Left panel: 1D regime $\rho(x,t)$; central panel: the profile $\rho(x,y,t_{last})$ taken as a snapshot at the last moment $t_{last}=1000$ of the modeled evolution; right panel: the time-sliced density plot of $\rho(x,y,t_n)$ at different moments $t_n$.	
}
\label{fig:4}
\end{figure}

Now we open a possibility for space-inhomogeneous solutions of Equations 
\eqref{psi} and \eqref{Gamma} considering one- and two-dimensional systems. 
The dynamical interplay of various fields gives rise to transient effects like the
self--trapping of excitons and the consequent nucleation of regions with
increased exciton density. The effect of self-trapping of excitons results in
the possibility of local phase transformations above a threshold
$\rho_{st}<\rho_{b}$ which is still lower than the barrier threshold $\rho_{b}$
for the homogeneous regime. The reason is the collective self--trapping
\cite{Rashba:88} of excitons similar to self--focusing in optics. 
For the intermediate initial pumping $\rho_{st}<\rho_{ini}<\rho_{b}$, the
self-trapping results in a dynamical formation of domains with different states of $\Psi$. 
Domains grow with time, they experience reflections from the boundaries, collisions and merging. That is accompanied by formation and
collapse of domain and domain walls and appearance of propagating wave fronts.
The numerical results for a sub-barrier pumping are presented in Figure \ref{fig:4}. 
 Its left panel shows an example for a 1D regime $\rho(x,t)$; this picture accompanies numerous illustrations which have been already presented in \cite{SB+NK:2016-prb}. The central and the left panels show results for a 2D system modeled upon a circle with homogeneous initial conditions. The central panel shows the three-domain profile $\rho(x,y,t_{last})$ taken as a snapshot at the last moment $t_{last}$ of the modeled evolution. The right panel shows this evolution in the form of the time-sliced density plot of $\rho(x,y,t_n)$ at different moments $t_n$. We see a roots-like picture where many domains are formed initially to experience a sequence of mergings till the stationary three-domains state.

 Future modeling for higher space dimensions will be of a certain principle interest in view of contradictory expectations in theory of dynamical solitons in dimensions $D\ge3$. On one side, the Derrick theorem claims an instability of stationary solitons; on another side the NLSE shows stable solitons as oscillatory solutions \cite{Vakhitov}.  The last case is closer to our model in view of intrinsically oscillating nature of the EI with allowed pairs creation.

\section{Excitonic Insulator, Multi--Field Model\label{Ch.EI-MF}}

At appearance of another, symmetry breaking, order parameters which we
associate with lattice displacements $u$, the energy functional acquires
additional terms $W(\Psi)\Rightarrow W(\Psi,u)=W(\Psi)+\Delta W(\Psi,u)$ where
\begin{align}
\Delta W(\Psi,u)  &  =
\Delta W(\rho,u)+\frac{\beta}{2}\left(\frac{\partial u}{\partial x}\right)^{2}
\label{W-psi-u}\\
\Delta W(\rho,u) & =\frac{d}{2}(\rho_{u}-\rho)u^{2}+\frac{f}{4}u^{4}
\end{align}
\noindent
Here $\rho_{u}$ is the threshold value of $\rho$ for appearance of
displacements $u$. The time evolution of $u$ is governed 
by the supplementary equation

\begin{equation}
K(\partial_{t}^{2}/\omega_{u}^{2}+\gamma\partial_{t})u=-\frac{\delta W(\Psi,u)}{\delta u}
\label{u-t} 
\end{equation}

\begin{figure}
\includegraphics[width=5cm]{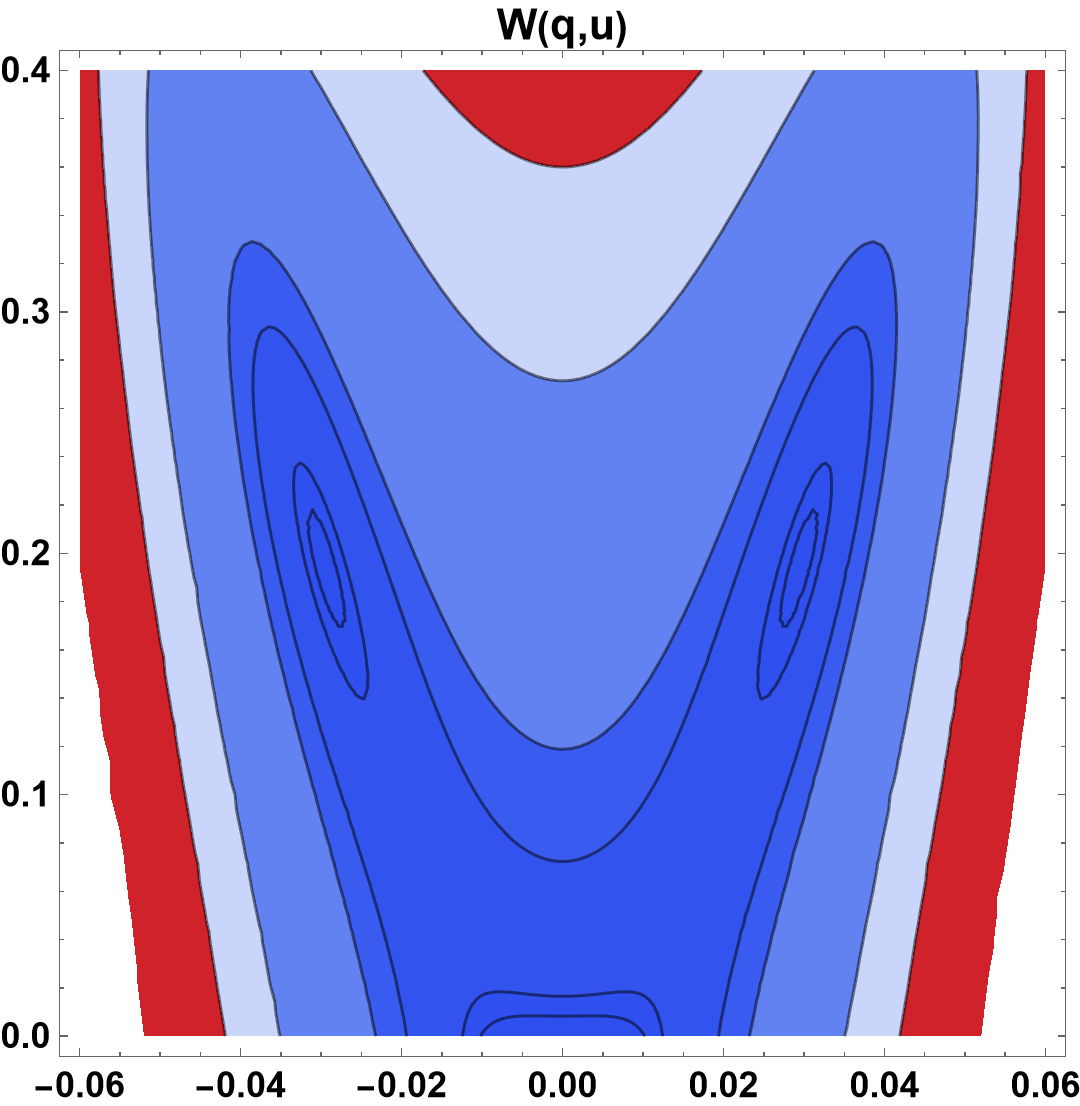}
\includegraphics[width=1cm]{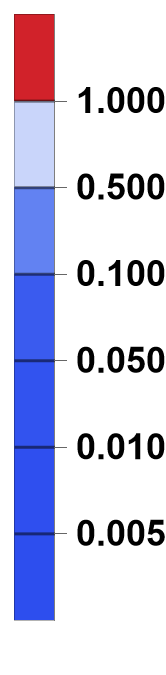}
\caption{{Density plot} of $W(\rho,u)$ for $\rho_{u}>0$. The double minima in $u$ are formed at the level $q\approx 0.2$.}
\label{fig:5}
\end{figure}

Near the BEC-EI crossover, the variable $\Psi$, hence $\rho$, is slow in
comparison with $u$ which is governed by the phonon frequency $\omega_{u}$.
Then it is helpful to draw the effective potentials 
$W^{\ast}(\rho)=W(\rho,u_{eq}(\rho))$ and $U^{\ast}(\rho)=U(\rho,u_{eq}(\rho))$ for the
equilibrium $u_{eq}(\rho)$ taken after the minimization of $W(\rho,u)$ over
$u$ at a given $\rho$. The plots for $W^{\ast}(\rho)$ and~$U^{\ast}(\rho)$ are mostly qualitatively similar to the plots for $W(\rho)$ and~$U^(\rho)$ in Figure~\ref{fig:1} for the generic EI. The difference is that now the energy function $W^{\ast}(\rho)$ demonstrates a new convex  region at small $\rho<\rho_{u}$, where the exciton energy $E_{ex}$ is
positive and it increases with $\rho$ corresponding to the repulsion of
excitons. The convex energy curve indicates that now there is no self-focusing at
small $\rho$, contrary to the situation for the generic EI (without lattice
deformations), where the concave curve starts from lowest $\rho$ (see Figure \ref{fig:1}).
The landscape of $W(\rho,u)$ for the full two-field model is presented in
Figure~\ref{fig:5}, to be compared with figure 6 in ref. \cite{crystals:17}.

At $\rho_{u}<0$, only the trivial minimum at $\rho=0,u=0$ does exist. 
At $\rho_{u}>0$ the double metastable minimum appear at $u=\pm u_{0}=\pm\sqrt{(\rho_{u}-\rho)d/f}$. 
At a higher $\rho_u>\rho_{uc}$ the double minima becomes stable while the trivial one
$(\rho=0,u=0)$ persists as a metastable one.

\subsection*{Time Evolution from the Numerical Modeling}

\begin{figure}
\includegraphics[width=0.4\linewidth]{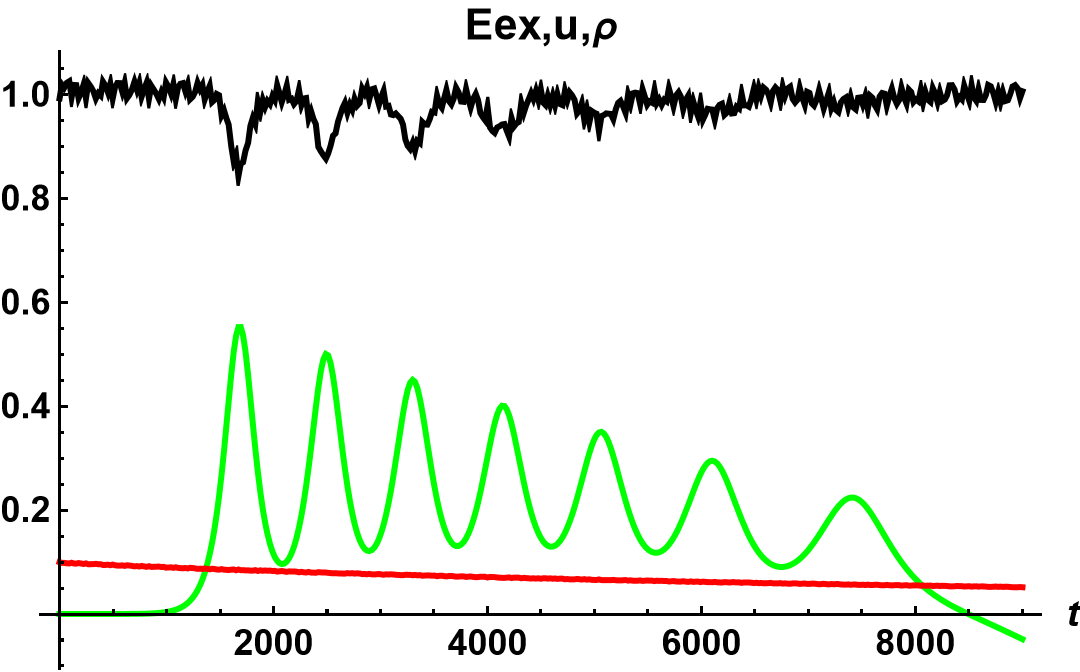}
\qquad
\includegraphics[width=0.4\linewidth]{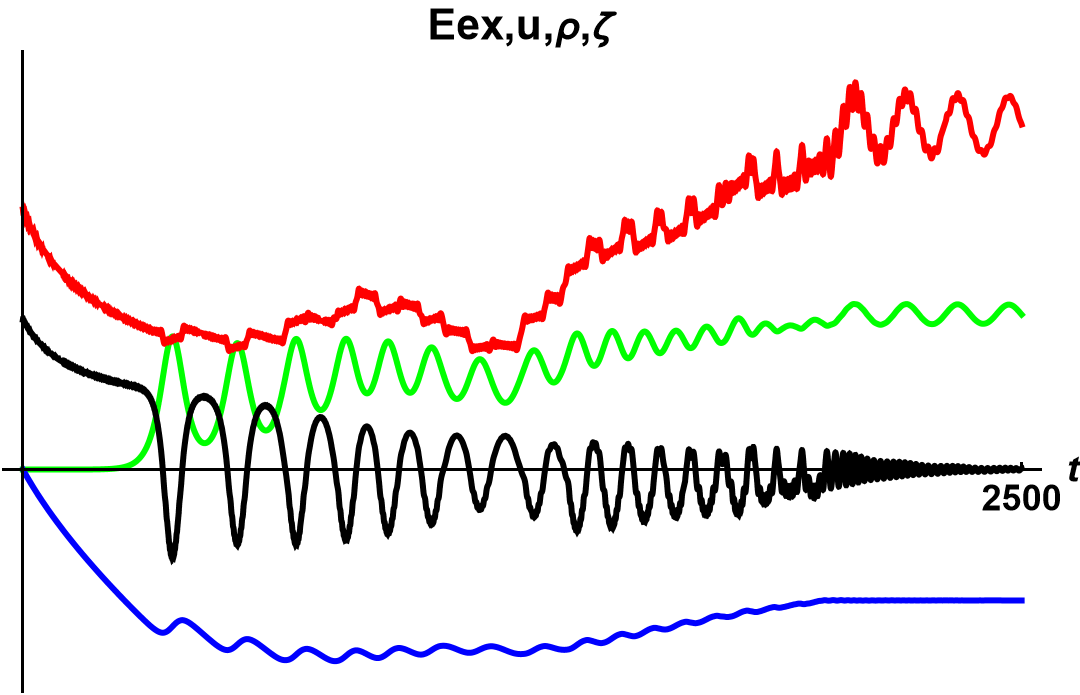}
\caption{ {Plots for $\rho(t)$ -- red, $u(t)$ -- green, and $E_{ex}(t)$ -- black. 
The left panels show results for a nearly subcritical pumping. The right panel shows results for a just supercritical pumping, with the added plot for the phase (in blue, rescaled).}}
\label{fig:6}
\end{figure}

	For the combined model with two order parameters we shall present results of
modeling only for homogeneous regimes (Figure~\ref{fig:6}). For the subcritical pumping, $u(t)$ temporarily emerges from the initial $u_{ini}=0$, then at large $t$ it vanishes together with $\rho\rightarrow 0$ while $E_{ex}$ stays nearly constant close to the unperturbed value. For the supercritical pumping both $\rho$ and $u$ reach, at long $t$, the new nontrivial equilibrium with vanishing exciton energy and the locked phase.
All quantities show two types of oscillations: common ones from the quantum interference and the specific ones with the frequency $\omega_{u}$ of the phonon mode in (\ref{u-t}).

\section{Effects of Hot Excitons}
\label{Ch.Incoh}

Above, we have assumed that a quasi--condensate of optically pumped excitons
appears sufficiently early as a macroscopic quantum state. 
It is desirable to take into account also the normal, non condensed density of excitons and its
(re)conversion (from)to the condensate. Indeed, temperatures in the
experiments are comparable with the estimated degeneracy temperature of the
BEC, and they are even higher just after the pumping pulse. The
not--quite--resonance pumping also contributes to the initial incoherent
density. Within this phenomenologically oriented article, we stay away from
the very important question of kinetics of non-condensed particles, relying on
experimental facts of the fast initial equilibration typical for PIPT.

A future microscopical study can be advanced thanks to the progress in theory
of equilibration in a gas of excitons and polaritons
\cite{Wouters:2007,Szymanska:2006,Gardiner:2002}, and to the general
understanding of a non equilibrium Bose gas motivated by problems in cold
atoms \cite{Kagan:93,Kagan:97}. For the ideal model of the weakly interacting
Bose gas, there is a fair overlap between the regime of the microscopical
kinetic and the collective NLSE--based descriptions. But the price is that a
turbulent mixing must be taken into account in the NLSE
\cite{Berloff:2002,Semikoz:97} and the Gross--Pitaevskii equation must be
considered stochastically rather than deterministically \cite{Gardiner:2002}.
For applications in solid state physics, the universal Bose gas model is not quite helpful as regards the BEC of excitons, because other channels of the relaxation like the emission of phonons \cite{Tikhodeev:90} become more
important than collisions of bosons.

To elucidate these effects we shall present below two types of modeling: 
i. stochastic equations taking into account the particles falling to the
condensate as a random noise; ii. phenomenological description for a dynamical
equilibrium with explicit exchange among the condensate and the incoherent reservoir.

\subsection{Stochastic Modeling at $d=0$}

The conversion of particles from the normal reservoir to the condensate  and the
related decoherence can be modeled numerically for a zero-dimensional system.
We need to generalize the NLSE to a stochastic differential equation.
The type of the random process must be chosen to simulate the random acts of
absorption of non-condensed particle into the condensate. The random function
$\nu$ needs to be added to the equation for the condensate density $\rho$.
That does not affect directly the equation for the phase, but the final effect
of the randomness upon the phase will be the most profound as it is
demonstrated by the numerical solution presented below.

In calculations, we used the Ito process:
\begin{equation}
d\rho=(-G\rho^{2}d\zeta+A\rho\sin(2\zeta))dt-d\nu~,~d\zeta
=(-U(\rho)+A\cos(2\zeta))dt
\end{equation}

The random function $\nu$ describes in average the exponential exhaustion of
the reservoir of the normal particles, emulating their diffusion in energy
space down to the condensate.

\begin{figure}
\includegraphics[width=7.5cm]{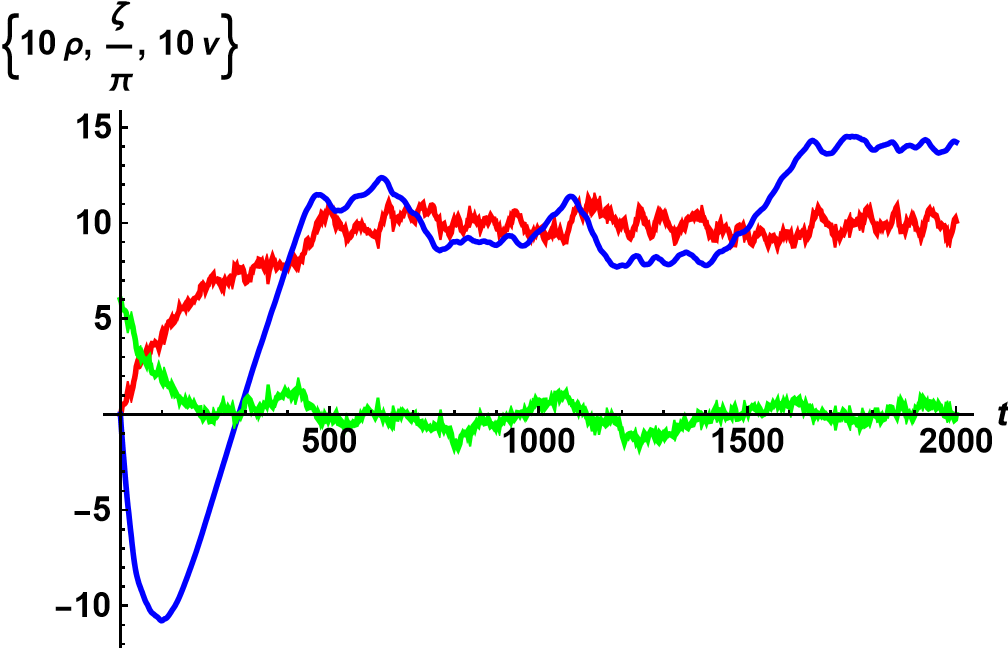}
\caption{{Plots of  $\nu(t)\times10$ -- green, $\rho(t)\times10$} -- red, $\zeta(t)/\pi$ for a supercritical pumping.}
\label{fig:7}
\end{figure}

Figure \ref{fig:7} presents the plots for the phase and the density of the
condensate and for the corresponding realization of the random process (chosen as
the exponential Brownian motion). The random increments of $\rho$ are
transferred to phase kicks, hence to the decoherence. Another effect, seen at
higher $T$ is the sequence of lock-in attempts with switchings among different
lock-in values of the phase.

\subsection{The Account for Incoherent Excitons and a Two--Fluid Description}
\label{2fluids}

Hidden problem in experimental feasibility of the BEC-EI scenario is the
transformation of the initially incoherent ensemble of pumped excitons to
their coherent Bose condensate. Initially hot excitons are expected to form a
conventional static ensemble of Bose particles, which occupation numbers of
normal states $|k>$ are $n_{k}$$\sim$$1$ with respect to the total large number of
particles; their wave functions $\psi_{k}\varpropto\exp(ikx-E_{k}t)$. Being
allowed to be quasi-static, which requires for some bottlenecks, this ensemble
can be described by Boltzmann kinetics. A more complicated quasi-stationary
version is a diffusive energy flow towards low energies \cite{Tikhodeev:90}. 
Under this cooling process, an intermediate state of a quasi-condensate is formed where at lowest
$E_{k}$ the occupation numbers start to be large $n_{k}>>1$. This part of the
ensemble can be described by a quasi-classical Gross-Pitaevskii equation,
while still in the turbulent regime, for the condensate with the kinetics for the
down-flow of normal particles. At the final equilibration, the occupation
numbers of lowest eigenstates $n_{0}$$\sim$$N$ and they can be described by the
single wave function augmented by the equilibrium Bose-Einstein distribution for normal particles.

The theory is simplified by considering the normal component as a separate
quasi--equilibrium reservoir, which can be characterized by the density $n$ or
the chemical potential $\mu_{n}$. This approach is validated by particular
spectra of polaritons, but it still needs to be justified for our system; we
briefly outline its possible application below as an absolutely minimalistic
description. A certain ground for the separation into two distinct,
particle--exchanging reservoirs comes from suggesting a bottleneck -- a
minimum $E_{min}$ of the kinetic energy -- where the pumped excitations are
accumulated after the initial rapid cooling. It is tempting to associate
$E_{min}$ with the energy of the lowest lattice mode interacting with
excitons. In our case, a good candidate is the soft mode in the dip of the
Kohn anomaly, which should exist as a precursor for the lattice dimerization
instability, (see \cite{DAvino} and the references therein). That can also be
the Debye frequency of the acoustical spectrum; both candidates converge to
$E_{min} \sim 100$ K.

We can make a simplifying, and quite plausible, suggestion that all reservoirs
of excitons contribute additively to the order parameter: the total charge
transfer becomes $\rho_{tot}=\rho+n$, where still $\rho=|\Psi|^{2}$. Then the
system energy and the particle potential become simply $W(\rho+n)$ and
$U(\rho+n)$ adding $\mu_{n}=U(|\Psi|^{2}+n)+E_{min}$. Now the Equation
(\ref{psi}) is further generalized to
\[
i\hbar\partial_{t}\Psi=
-\frac{\hbar^{2}}{2M}\partial_{x}^{2}\Psi+U(|\Psi|^{2}+n)\Psi+\frac{i}{2}Rn\Psi-\frac{A}{2}\Psi^{\ast}
\]
where $R$ is a conversion rate regulating the exchange among the reservoirs.
This equation needs to be complemented by an equation for $n$, which we choose
as a simple rate equation with diffusion (cf. \cite{Wouters:2007})
\[
\partial_{t}n-\partial_{x}b\partial_{x}\mu_{n}n=I-Rn|\Psi|^{2}
~,~|\Psi|^{2}=\rho
\]
where $\mu_{n}$ and $b$ are the chemical potential and the mobility of
normal particles. $I(t)$ is the pump intensity profile; being short, it can be
omitted in favor of the initial condition $n(0)=n_{0}=\int I(t)dt$. The
function $R$ must change the sign as a function of the discrepancy $\delta
\mu=\mu_{n}-\mu_{c}$ of chemical potentials in the normal and the condensed
subsystems. We shall adopt for $R$ the simplest linear form which is valid close to
equilibrium at $|\delta\mu|\ll T$; otherwise, it can be generalized to
$R\propto\sinh(\delta\mu/T)$ or to a more complicated non symmetric form. With
a common definition for the chemical potential $\mu_{c}$ of the BEC, we have
\[
\mu_{c}=-\hbar\partial_{t}\zeta+\frac{\hbar^{2}}{2M}(\partial_{x}\zeta)^{2}
~,~\mu_{n}=U(|\Psi|^{2}+n)+E_{min}~,~R=kn(\mu_{n}-\mu_{c})/\hbar,
\]

The simpler space--independent Equations \eqref{dphase-dt} and \eqref{drho-dt} are
generalized as
\begin{align}
\hbar\dot{\zeta} &  =-U(\rho+n)+A\cos(2\zeta),\\
\dot{\rho} &  =\rho R+A\rho\sin(2\zeta)=
k\rho n(\dot{\zeta}+\mu_{n}/\hbar)+A\rho\sin(2\zeta),
\nonumber\\
&  =(k/\hbar)\rho n(E_{min}+A\cos(2\zeta))+A\rho\sin(2\zeta)\\
\dot{n} &  =I-k\rho n(\dot{\zeta}+\mu_{n}/\hbar)
=I-(k/\hbar)\rho n(E_{min}+A\cos(2\zeta)).
\end{align}

The results for the case of pumping to the reservoir of uncondensed excitons have been presented above in plots of Figure  \ref{fig:3}.
At the subcritical pumping $n(0)=0.27$ all quantities show
oscillations which are particularly pronounced in $E_{ex}(t)$. The phase
decreases monotonically, correspondingly $E_{ex}(t)>0$, hence the regime of
the BEC. The concentration of condensed particles $\rho(t)$ initially grows, being
fed by the decreasing population $n(t)$ of the normal reservoir, then it
passes through the maximum and vanishes at long time where $E_{ex}(t)\rightarrow E_{ex}^{0}(t)$. 
The polar plot shows only one trivial circulation around $\rho=0$.

For the supercritical pumping $n(0)=0.28$ the phase start to
decrease, correspondingly $E_{ex}(t)>0$ for this initial regime of the BEC.
Passing through the minimum, the phase starts to grow, correspondingly
$E_{ex}(t)<0$ indicating the regime, still nonstationary, of the EI. Finally
the phase is locked at the value $\zeta(t)\rightarrow\pi$. The number of
condensed particles $\rho(t)$ shows an average growth superimposed by regular
oscillations; it keeps growing even after the population $n(t)$ of the normal
reservoir is exhausted; later the growth saturates at the equilibrium position
$\rho_{st}\approx1$ of the static EI state. $E_{ex}(t)$ shows the most
irregular behavior to finally saturate at $E_{ex}(t)\rightarrow0$ as it should
be for the static EI state. The polar plot shows two circulations: first
around $\rho=0$, then switching to the one around $\rho=1$.

For both cases we have presented the results very close to the dynamical
pumping threshold which happens sharply between $n_{0}=0.27$ and $n_{0}=0.28$
- well below the thermodynamical threshold at $\rho_{0}=0.36$.

\section{Conclusions}

\label{Ch.Con}
The concepts and methods of pump-induced phase transitions can be applied to non-metallic systems where ubiquitous  optical excitons can play a role of pumping agents, unlike the most common practice of pumping to unbound e-h pairs. If the phase diagram of studied material shows a multitude of stable and metastable states, then the pumping can achieve switching among these states or even create new static or stationary ones. In general, the excitons and the thermodynamical order parameter can be independent while mutually affecting as we have briefly outlined in Ch.2. But mostly we payed attention to a more specific class which is of a particular interest for theory and of strong experimental implications. Experimentally, this is the case of neutral-ionic transitions in donor-acceptor structures which allow for transformations by both thermodynamical means and the optical pumping. Specifically, here the excitation and the long range ordering are built from the same intermolecular electronic transfers, so that the density of pumped excitons contributes additively to the thermodynamic order.
Both thermodynamical and dynamical effects can be described on the same root by viewing the ordered state as the Excitonic Insulator which appears as a
macroscopic quantum state of excitons starting as their Bose condensate allowed at high pumping densities.
The double nature of the ensemble of excitons leads to an intricate time evolution: the dynamical transition between number–preserved and phase–locked regimes, macroscopic quantum oscillations from interference between the Bose condensate of excitons and the ground state of the excitonic insulator. Modeling of an extended sample shows also stratification in domains of low and high densities which evolve through local dynamical phase transitions and a sequence of domains’ merging. The locally enhanced density of excitons can surpass a critical value to trigger the local persisting phase transformation, even if the mean density is below the required threshold.

Beyond our phenomenological descriptions, there are challenging questions of the relaxations of initial hot reservoir of excitons to the growing condensate and of the establishing of the coherence. The microscopic aspects of these difficult questions have not been completely resolved even in much simpler systems like cold atoms or polaritons. Here we have suggested two kinds of illustrations. One is the stochastic generalization of the macroscopic equations to emulate random absorption of incoherent excitons to the condensate. Another illustration is the rate equation regulating exchange of normal and condensed reservoirs controlled by a mismatch of chemical potentials.

We believe that the presented study and discussions will attract attention to still ill explored excitonic routes in the PIPT science.

\end{document}